\input harvmac
\def\half{{1 \over 2}}

\def\p{{\partial}}
\def\s{{\sigma}}

\def\a {{\alpha}}
\def\b {{\beta}}

\def\g {{\gamma}}

\def\ad {{\dot\alpha}}
\def\bd {{\dot\beta}}

\def \ad {{\dot \a}}
\def \bd {{\dot \b}}

\Title{\vbox{\hbox{IFUSP-P-1213}}}
{\vbox{\centerline{\bf Off-shell 
Supersymmetry versus Hermiticity in the
Superstring}}}
\bigskip\centerline{Nathan Berkovits}
\bigskip\centerline{Dept. de F\'{\i}sica Matem\'atica, Univ. de S\~ao Paulo}
\centerline{CP 20516, S\~ao Paulo, SP 01498, BRASIL}
\centerline{and}
\centerline{IMECC, Univ. de Campinas}
\centerline{CP 1170, Campinas, SP 13100, BRASIL}
\bigskip\centerline{e-mail: nberkovi@snfma2.if.usp.br}
\vskip .2in

We point out that off-shell four-dimensional spacetime-supersymmetry implies
strange hermiticity properties for the N=1 RNS superstring. However, these
hermiticity properties become natural when the N=1 superstring is embedded
into an N=2 superstring.

\Date{April 1996}

In four-dimensional compactifications of the N=1 RNS superstring, the
spacetime-supersymmetry generators in the $-\half$ picture are
\eqn\one{q_a={1\over{2\pi i}}
\oint dz
~ e^{{1\over 2}(-\phi \pm i\s_0\pm i\s_1\pm iH_C)}}
where there are an even number of $+$ signs in the exponential
($a=1$ to 4), $\phi$ comes from fermionizing the bosonic ghosts
as $\b=i\p\xi e^{-\phi}$ and $\g=-i\eta e^\phi$,
$\psi^3\pm\psi^0=e^{\pm i\s_0}$ and
$\psi^1\pm i\psi^2=e^{\pm i\s_1}$
where $\psi^m$ is the fermionic vector, and $\p H_C=J_C$ is
the U(1) generator of 
the c=9 N=2 superconformal field theory representing
the compactification manifold.

These spacetime-supersymmetry generators
satisfy the anti-commutation relations
\eqn\two{\{ q_a, q_b\}={1\over{2\pi i}}
\oint
dz~ e^{-\phi} \psi_m \gamma^m_{ab}}
which is not the usual supersymmetry algebra
$\{ q_a, q_b\}={1\over{2\pi }}
\oint dz~ \p x_m \gamma^m_{ab}$ where
${1\over {2\pi}}\oint dz~ \p x_m$ is the string
momentum.
However after hitting the right-hand side of \two with the picture-changing
operator $Z=\{Q,\xi\}$, it becomes
${1\over{2\pi i}}
\oint dz~ Z e^{-\phi} \psi_m \gamma^m_{ab}={1\over{2\pi }}
\oint dz~
\p x_m \gamma^m_{ab}.$
So up to picture-changing, the $q_a$'s form a supersymmetry algebra.\ref
\friedan{D. Friedan, E. Martinec and S. Shenker, Nucl. Phys. B271
(1986) p.93.}

But off-shell supersymmetry requires that
the $q_a$'s 
form a supersymmetry algebra without applying
picture-changing operations. 
This is because picture-changing is only well-defined when the
states are on-shell. Off-shell, the states are not independent
of the locations of the picture-changing operators. 

So off-shell spacetime-supersymmetry requires modification of the
$q_a$'s. Note that $q_a$ has picture $-\half$ and the momentum
${1\over{2\pi }}
\oint dz~\p x_m $
has picture 0, so we need generators with picture $+\half$.
The obvious solution\ref\four
{N. Berkovits, Nucl. Phys. B431 (1994) p.258.}
is to split $q_a$ into a chiral part with
picture $-\half$ and an
anti-chiral part with picture $+\half$:
\eqn\three{q_\a= 
{1\over{2\pi i}}
\oint dz~  e^{{1\over 2}(-\phi\pm i(\s_0 +\s_1)+ iH_C)}}
$$\bar q_\ad=Z q_\ad=
{1\over{2\pi i}}
\oint dz~ [b\eta e^{{1\over 2}(3\phi\pm i(\s_0-\s_1)- iH_C)}
$$
$$+i ~:( e^\phi\psi_m\p x^m + e^\phi G_C^+ +
e^\phi G_C^-) e^{{1 \over 2}(-\phi\pm i(\s_0-\s_1)- iH_C)}:]$$
where $G_C^\pm$ are the fermionic generators of the 
c=9 N=2 superconformal field theory.
The N=1 4D supersymmetry algebra
$\{q_\a, \bar q_\bd\}=
{1\over{2\pi }}
\oint dz~\p x_m \s^m_{\a\bd}$ 
is now satisfied
off-shell where we are using standard two-component Weyl notation.

Although we have solved the problem of finding off-shell
supersymmetry generators, we now have a new problem. 
Using the standard RNS
definition of hermiticity where all fundamental fields are hermitian
or anti-hermitian (the anti-hermitian field is $\s_0$),
the hermitian conjugate of $q_\a$
is no longer $\bar q_\ad$. Fortunately, this new problem can
be solved by modifying the definition of hermiticity. 
However, this new hermiticity definition will only be natural if one
embeds the N=1 superstring into an N=2 superstring.

To find the appropriate hermiticity definition, one
first writes $\bar q_\ad$ in the form
$$\bar q_\ad = e^R~({1\over{2\pi i}}
\oint dz~ b \eta e^{{1\over 2}
{(3\phi \pm i(\s_0 -\s_1)-iH_C)}} )~e^{-R}$$
where 
\eqn\R{R={1\over{2\pi }}
\oint dz~ c \xi e^{-\phi} (\psi^m \p x_m +G^+_C+G^-_C)}
and $e^R F e^{-R}= F +[R,F] +\half [R,[R,F]] + ...$ (the expansion usually
stops after two terms).

One then defines hermiticity as:
\eqn\herm{(x_m)^\dagger = e^R x_m e^{-R},\quad
(\psi_m)^\dagger = e^R \psi_m e^{-R},\quad,
(F_C)^\dagger = e^R \bar F_C e^{-R},}
$$(e^{\phi\over 2})^\dagger = e^R( c \xi e^{-{3\over 2}\phi}) e^{-R},\quad
(e^{-{\phi\over 2}})^\dagger = e^R( b \eta e^{{3\over 2}\phi}) e^{-R},$$
$$(b)^\dagger = e^R (i b \eta \p\eta e^{2\phi}) e^{-R},\quad
(c)^\dagger = e^R (-i c \xi \p\xi e^{-2\phi}) e^{-R},$$
$$(\eta)^\dagger = e^R(i \eta b \p b e^{2\phi}) e^{-R},\quad
(\xi)^\dagger = e^R (-i\xi c \p c e^{-2\phi}) e^{-R},$$
where $F_C$ are the worldsheet fields in the c=9 N=2 superconformal
field theory.
It is straightforward to check that the new
hermiticity definition satisfies $(F^\dagger)^\dagger =F$ for
all $F$, preserves 
OPE's,
and implies that $(q_\a)^\dagger = \bar q_\ad $. 

One strange feature of the hermiticity definition of \herm 
is that a field may have a different conformal weight
from its hermitian conjugate since $(T)^\dagger= T+
i\p (bc +\xi\eta)$ where $T$ is the RNS Virasoro generator. 
Another strange feature is that the BRST operator
is not hermitian since
$Q^\dagger={1\over{2\pi }}
\oint dz~b.$ (This easily follows from writing
$Q=e^R({i\over{2\pi }}
\oint dz~ b\eta\p\eta e^{2\phi})e^{-R}$.)

Although these features are strange in the N=1 RNS description of the
superstring, they are natural if the N=1 superstring is embedded into
an N=2 superstring. As discussed in reference \ref\ust
{N. Berkovits and C. Vafa, Mod. Phys. Lett. A9 (1994) p.653.},
any critical N=1 superstring can
be embedded into a critical N=2 superstring where
the c=6 N=2 superconformal generators are\ust
\eqn\gentwo{T_{N=2} =T_{N=1} +{i\over 2}
\p(bc+\xi\eta),\quad G^+_{N=2}=j_{BRST},\quad
G^-_{N=2}=b,\quad J_{N=2}= bc+\xi\eta,}
and $j_{BRST}=e^R (i b\eta\p\eta e^{2\phi}) e^{-R}$.

Using the hermiticity definition of \herm, $(T_{N=2})^\dagger=T_{N=2}$,
$(G^+_{N=2})^\dagger=G^-_{N=2}$, and $(J_{N=2})^\dagger=J_{N=2}$,
which are the standard hermiticity properties of an N=2 string. (This
hermiticity can be made manifest by writing the N=2 generators of \gentwo
in terms of spacetime-supersymmetric variables.\four)
So the hermiticity properties implied by off-shell four-dimensional
supersymmetry are natural only if the N=1 superstring is embedded into
an N=2 superstring.
\listrefs
\end